\documentstyle[12pt]{article}
\begin{document}
\begin{center}
{\large\bf A HEURISTIC REMARK \\
ON THE PERIODIC VARIATION\\
IN THE NUMBER OF SOLAR NEUTRINOS\\
DETECTED ON EARTH}\\ [0,5cm]
H.J. Haubold\\[0,3cm]
UN Outer Space Office, Vienna International Centre, Vienna,
Austria\\[0,5cm]
and\\[0,5cm]
A.M. Mathai\\[0,3cm]
Department of Mathematics and Statistics, McGill University,
Montreal, Canada\\
\end{center}
\noindent
Abstract. Four operating neutrino observatories confirm the long
standing discrepancy between detected and predicted solar
neutrino
flux. Among these four experiments the Homestake experiment is
taking data for almost 25 years. The reliability of the
radiochemical method for detecting solar neutrinos has been
tested
recently by the GALLEX experiment. All efforts to solve the solar
neutrino problem by improving solar, nuclear, and neutrino
physics
have failed so far. This may also mean that the average solar
neutrino flux extracted from the four experiments may not be the
proper quantity to explain the production of neutrinos in the
deep
interior of the Sun. Occasionally it has been emphasized that the
solar neutrino flux may vary over time. In this paper we do
address
relations among specific neutrino fluxes produced in the
proton-proton chain that are imposed by the coupled systems of
nonlinear partial differential equations of solar structure and
kinetic equations by focusing our attention on a statistical
interpretation of selected kinetic equation of PPII/PPIII branch
reactions of the proton-proton chain. A fresh look at the
statistical implications for the outcome of kinetic equations for
nuclear reactions may shed light on recent claims that the
$^7Be$-neutrino flux of the Sun is suppressed in comparison to
the
pp- and $^8$B neutrino fluxes and may hint at that the solar
neutrino flux is indeed varying over time as shown by the
Homestake
experiment.
\clearpage

\section{Solar Nuclear Energy Generation: Proton-Proton-Chain}
The nuclear energy source in the Sun is believed to be the
proton-proton chain, in which four protons fuse to form one
$^4He$
nucleus,
i.e.
\begin{equation}
4p\rightarrow  ^4He+2e^++2\nu_e +Q,
\end{equation}
where $Q=M(^4He)-4M_p-2M_e\approx 26.73 MeV$ denotes the energy
release. The three different branches (PPI, PPII, PPIII) to
accomplish the
formation of $^4He$ in the pp-chain are shown in Figure 1.
Neutrinos are produced in the  pp-chain by nuclear fusion
reactions,  beta-decay, and  electron capture.
The dominant reactions (PPI, 86\% of the produced $^4He$) produce
the great majority of low energy solar neutrinos
$(\Phi_\nu^{SSM}(pp)\approx 6.0 \times 10^{10}\nu
cm^{-2}s^{-1}, \Phi^{SSM}_\nu(pp)\sim T_c^{-1.2}$, where $T_c$
denotes the temperature at the centre of the Sun) and their
number
should be a firm prediction of any solar model because it is
closely tied
to the solar luminosity. The second branch of the pp-chain
(PPII, 14\% of the produced $^4He$) yields the $^7Be$ neutrinos
at
two discrete energies
$(\Phi ^{SSM}_\nu(^7Be)\approx 4.9\times 10^9 \nu cm^{-2}s^{-1},
\Phi^{SSM}_\nu(^7Be)\sim T_c^8).$ In the third branch of the
pp-chain, the very rare case (PPIII, 0.02\% of the produced
$^4He$), radioactive $^8B$ is produced which decays ultimately
and
is the source of high energy $^8B$ neutrinos $(\Phi^{SSM}_\nu
(^8B)\approx 5.5 \times 10^6\nu cm^{-2}s^{-1},
\Phi_\nu^{SSM}(^8B)\sim T_c^{18}).$ The specific neutrino fluxes
are taken from the Standard Solar Model (SSM) of Bahcall and
Pinsonneault (1992). The overall energy production of the
pp-chain
is $Q=26.73 MeV$, however, the three branches produce a different
amount of energy (PPI: $E_\gamma=(26.73-0.265)$MeV, PPII:
$E_\gamma=(26.73-0.861)$MeV, PPIII: $E_\gamma=(26.73-7)MeV)$ due
to
the energy loss carried off by elusive neutrinos. The fluxes of
high energy solar neutrinos are especially sensitive to the
central
temperature, $T_c$, mainly because of the energy dependence of
the
cross sections of the respective nuclear reactions. The branching
ratios (the percentage of which each branch of the pp-chain
contributes to the production of $^4He$) are strongly dependent
on
the nuclear reaction probabilities and on the density and
temperature profiles inside the Sun. Assuming that the Sun is in
a
state of quasistatic equilibrium, the solar luminosity $L_\odot$
tells us the total energy generation rate which can be turned
into
a constraint on the total solar neutrino fluxes, that is
\begin{equation}
L_\odot=13.1(\phi_\nu(pp)-\phi_\nu(^7Be)-\phi_\nu(^8B))+25.6\phi_
\nu(^7Be)+19.5\phi_\nu(^8B).
\end{equation}
The luminosity $L_\odot$ observed at the current stage of
evolution
of the Sun corresponds to the energy that was generated in the
gravitationally stabilized solar fusion reactor $10^7$ yr  ago
(Helmholtz-Kelvin timescale). The quasistatic assumption allows
one
to equate the present luminosity with the present nuclear energy
production rate (Mathai and Haubold, 1988).
Normalizing the neutrino fluxes in (2) to those of the Standard
Solar Model $(\Phi_\nu=\phi_\nu/\phi_\nu^{SSM})$ leads to a
luminosity constraint indicating the degree of contribution of
the
respective neutrino flux to the total solar neutrino emission:
\begin{equation}
1=0.913\Phi_\nu(pp)+0.071\Phi_\nu(^7Be)+0.00004 \Phi_\nu(^8B).
\end{equation}
To reveal how the pp-chain operates in the Sun is to measure the
individual neutrino fluxes $\phi_\nu(pp),\, \phi_\nu(^7Be),$ and
$\phi_\nu(^8B)$ in (3), thereby fixing the branching ratios of
PPI,
PPII, and PPIII as indicated in Figure 1. Four experiments are in
operation now to accomplish this solar neutrino spectroscopy.
Kamiokande measures exclusively the high-energy flux
$\phi_\nu(^8B)$, thus PPIII, in real-time spectroscopy. Homestake
observes primarily $\phi_\nu(^8B)$ and to a much lesser extend
$\phi_\nu(^7Be)$, that is, the branching of PPII and PPIII.
GransSasso/Baksan detects primarily the low-energy flux
$\phi_\nu(pp)$ and to a lesser extend the fluxes $\phi(^7Be)$ and
$\phi_\nu(^8B)$, thus focusing on the branching of PPI and
PPII/PPIII.

\section{Spatial Distribution of Solar Neutrino Sources: Standard
Solar Model}
Figure 2 shows the neutrino production as a function of the
dimensionless distance variable $x=R/R_\odot$, starting from the
center of the Sun for the Standard Solar Model (Bahcall and
Pinsonneault, 1992). We note that experiments for the detection
of
solar neutrinos are looking into different depths of the solar
core
as they are only sensitive to specific neutrinos produced in the
respective nuclear reactions. The region in which the low-energy
pp-neutrino flux is produced is very similar to that of the total
nuclear energy generation. Because of its strong temperature
dependence, the high-energy $^8B$-neutrino production is peaked
at
the very small radius $x=0.05$ and is generated in a much
narrower
region in comparison to the other two neutrino sources (Table
3).\par
The contribution of the neutrino fluxes $\phi_\nu(pp),
\phi_\nu(^7Be),\, \mbox{and}\, \phi_\nu(^8B)$ to the Homestake,
Kamiokande, and GranSasso/Baksan experiments is
\begin{equation}
\phi_\nu\mbox{(Homestake)}\approx
6.2\phi_\nu(^8B)+1.2\phi_\nu(^7Be) SNU,
\end{equation}
\begin{equation}
\phi_\nu\mbox{(Kamiokande)}=\phi_\nu(^8B)\Phi_\nu(^8B),
\end{equation}
\begin{equation}
\phi_\nu\mbox{(GranSasso/Baksan)}\approx
13.8\phi_\nu(^8B)+35.8\phi_\nu(^7Be)+70.8\phi_\nu(pp).
\end{equation}
The coefficients in Equation (4)-(6) are corresponding to the
capture rates predicted by the Standard Solar Model for each
respective solar neutrino source.\par
Table 1 summarizes the predictions of the total capture rates of
the Standard Solar Model for the Kamiokande, Homestake, and
GranSasso/Baksan experiments. The Kamiokande solar neutrino flux
is
given in units of\\
$10^6 \nu cm^{-2}s^{-1}$, while the Homestake and
GranSasso/Baksan
rates are given in solar neutrino units $(1SNU \equiv 10^{-36}\nu
atom^{-1} s^{-1}).$ The uncertainties shown in Table 1 are
$1\sigma$.\par
\bigskip
\begin{tabular}{|l|c|c|c|}\hline
Experiment & Kamiokande & Homestake & GranSasso/Baksan\\ \hline
Predicted & $5.69\pm 0.82$ & $8.0\pm 3.0$ & $131.5  \pm
^{21}_{17}$\\
capture rate & & & \\ \hline
\end{tabular}\par
\vspace{0.5cm}
Table 1.

\section{Four Solar Neutrino Experiments: Solar Neutrino Problem
and   $^7$Be-Neutrino Deficiency}
\begin{tabular}{|p{1.5in}|p{2.in}|c|l|}\hline
Experiment & Reaction & Energy & Location \\
& & threshold & \\ \hline
Kamiokande & $\nu_e+e^-\rightarrow \nu_e+e^-$ & $7.5 MeV$ & Japan
\\ \hline
Homestake & $\nu_e+^{37}Cl\rightarrow ^{37}Ar+e^-$ & $0.814 MeV$
&
USA\\ \hline
GranSasso & $\nu_e+^{71}Ga \rightarrow ^{71}Ga+e^-$ & $0.233 MeV$
& Italy\\
(GALLEX) & & &\\ \hline
Baksan & $\nu_e+^{71}Ga\rightarrow ^{71}Ge+e^-$ & $0.233 MeV$ &
Russia\\
(SAGE) & & &\\ \hline
\end{tabular}\par
\vspace{0.5cm}
Table 2.\par
There are four experiments currently operating to detect
neutrinos
coming from the Sun (Table 2). The Kamiokande experiment is a
water
\v{C}erenkov detector which measures the energy of the scattered
electrons (Nakamura, 1993). Due to its energy threshold it is
only
sensitive to the high-energy $^8B$ neutrinos from branch PPIII of
the pp-chain. The Homestake experiment consists of $10^5$ gallons
of $C_2Cl_4$ and detects solar neutrinos via capture on the
chlorine (Davis, 1993). Its energy threshold allows to detect the
higher energy line of $^7Be$-neutrinos from branch PPII as well
as
the high-energy $^8B$ neutrinos from branch PPIII. The two
gallium
experiments at GranSasso and Baksan are sensitive to the low
energy
pp-neutrinos from the PPI branch of the pp-chain as well as to
the
higher energy $^7Be$- and $^8B$-neutrinos (Anselmann et al.,
1994;
Abdurashitov et al., 1994). The predicted contributions to the
Homestake and GranSasso/Baksan experiments based on the Standard
Solar Model are shown in Table 3.\par
\bigskip
\begin{tabular}{|l|lcr|l|}\hline
Neutrino & Homestake & & Percentage & GranSasso/Baksan\\
source & experiment & & of total & experiments\\
& & & capture rate & \\ \hline
pp & $0.0$ & $\equiv$ & $0\%$ & $70.8$\\ \hline
pep & $0.2$ & $\equiv$ & $2.5\%$ & 3.1\\ \hline
$^7Be$ & $1.2$ & $\equiv$ & $15\%$ & $35.8$\\ \hline
$^8B$ & $6.2$ & $\equiv$ & $77.5 \%$ & $13.8$\\ \hline
$^{13}N$ & $0.1$ & $\equiv$ & $1.25\%$ & $3.0$\\ \hline
$^{15}O$ & $0.3$ & $\equiv$ & $3.75\%$ & $4.9$\\ \hline
total & $8.0\pm 3.0$ & $\equiv$ & $100\%$ & $131.5\pm
^{21}_{17}$\\

capture rate & & & & \\ \hline
\end{tabular}\par
\vspace{0.5cm}
Table 3.\par
\bigskip
The experimental results of the four solar neutrino experiments
are
given in Table 4 and can be compared with the predictions of the
Standard Solar Model as shown in Table 1.\par
\bigskip
\begin{tabular}{|l|c|l|}\hline
Experiment & Detected & Detected capture rate/\\
& capture rate & predicted capture rate\\ \hline
Kamiokande & $2.89^{+0.22}_{-0.21}\pm 0.35$ & $0.50 \pm 0.07$\\
\hline
Homestake & $2.55 \pm 0.17\pm 0.18$ & $0.32 \pm 0.03$\\ \hline
GranSasso/ & $77\pm9$ & $0.59 \pm 0.07$\\
Baksan & & \\ \hline
\end{tabular}\par
\vspace{0.5cm}
Table 4.\par
\medskip
\noindent
{}From Table 4 it is evident that the results of the four
experiments
are between 1/3 and 1/2 of the neutrino capture rates predicted
by
the Standard Solar Model. This deficit of solar neutrinos is
called
the solar neutrino problem which poses a serious conflict with
the
constraint of the overall solar luminosity in (2) and (3).
Additionally, the comparison of the three detected capture rates
in
Table 4 with the predicted capture rates in Tables 1 and 3 shows
that the Kamiokande rate is less suppressed than the Homstake
rate.
Because the Homestake experiment has a lower energy threshold,
the
lower detected capture rate suggests that the $^7Be$-neutrinos
are
more suppressed than the high energy $^8B$-neutrinos. However,
any
reduction of the $^7Be$ production rate by lowering the
temperature
$T_c$ would affect immediately both the $^7Be$ and $^8B$ neutrino
production equally. This fact seems to pose an additional problem
in finding a solution of the solar neutrino problem in terms of
solar, nuclear, and neutrino physics on which the Standard Solar
Model is based. This is particularly true for the so-called
cooler
Sun models (Castellani et al., 1994).

\section{Argon-Production Rate of the Homestake Experiment:
Variations Over Time}
Figure 3 shows the $^{37}Ar$ production rate detected by the
Homestake experiment from 1970.8 to 1991.6 (Davis, 1993). The
average $^{37}Ar$ production rate (combined likelihood function)
for the 94 individual runs shown was $0.509\pm 0.031$ argon atoms
per day. Subtracting a total background $^{37}Ar$ production rate
of $0.08\pm 0.03$ atoms per day yields the production rate that
can
be ascribed to solar neutrinos: $0.429\pm 0.043$ atoms per day or
$2.28\pm 0.23$ SNU (the rate in SNU is equal to $5.31$ times the
captures per day in the Homestake experiment). This average
capture
rate is commonly compared to the predictions of the Standard
Solar
Model as shown in Tables 1 and 3 for the pp-chain and the CNO
cycle. This procedure does not take into account in any way the
apparent time variation in the observed $^{37}Ar$ production rate
evident in Figure 3.\par
Figure 4 shows a five-point moving average of the $^{37}Ar$
production rate, removing high frequency noise from the actual
time
series collected in the Homestake experiment as shown in Figure
3.
One notes in the five-point moving average that in the periods
1978
to 1979 and 1987 to 1988 a supression of the $^{37}Ar$ production
rate seems to occur. The overall shape of the five-point moving
average suggests that there are two distinctive epochs spanning
the
time periods 1971 to 1980 and 1980 to 1989. Each epoch shows a
shock-like rise and subsequent rapid decline of the $^{37}Ar$
production rate. Further, the five-point moving average of the
$^{37}Ar$ production rate
reveals that each of the two distinct cycles covers a time period
of around 9 years. Each cycle exhibits a slow but shock-like
increase, reaching a peak, succeeded by a rapid decrease to a
minimum value of the $^{37}Ar$ production rate. This pattern is
repeated for a second nine-year period and seems to start for a
third period in 1989 (Haubold and Mathai, 1994). Each of these
cycles can be reproduced by a mechanism discussed in the
following
Section.
Fourier analysis of the $^{37}Ar$ production rate data in Figure
3
reveals a power spectrum showing the harmonic content in this
time
series in terms of a series of distinctive periodicities which is
shown in Figure 5.
Fourier analysis also indicates that the harmonic content in the
$^{37}Ar$ production rate data is dominated by periodicities of
0.57, 2.2, 4.8, and 8.3 years (Haubold and Gerth, 1990).

\section {Kinetic Equations: Lifetime Densities}
The production and destruction of nuclei in the proton-proton
chain
of reactions can be described by kinetic equations governing the
change of the number density $N_i$ of species $i$ over time, that
is,
\begin{equation}
\frac{d}{dt}N_i=-\sum_jN_iN_j<\sigma v>_{ij} + \sum_{k,l\neq
i}N_kN_l<\sigma v>_{kl},
\end{equation}
where $<\sigma v>_{mn}$ denotes the reaction probability for  an
interaction involving species $m$ and $n$, and the summation is
taken over all reactions which either produce or destroy the
species $i$. The first sum in (7) can also be written as
\begin{equation}
-\sum_jN_iN_j<\sigma v>_{ij}=-N_i(\sum_j N_j<\sigma
v>_{ij})=-N_ia_i,
\end{equation}
where $a_i$ is the statistical expected number of reactions per
unit volume per unit time destroying the species $i$. The
reciprocal of the quantity $a_i$ is the lifetime of species $i$
for
interaction with species $j$ for all $j$. It is also a measure of
the speed in which the reaction proceeds. If the reaction results
in the production of a neutrino, for example, then the reciprocal
of $a_i$ is the expected time it takes to produce this neutrino
in
the solar interior. In the following we are assuming that there
are
$N_j(j=1,\ldots, i, \ldots)$ of species $j$ per unit volume and
that for a fixed $N_i$ the numbers of other reacting species that
react with the i-th species are constants in a unit volume.
Following the same argument we have for the second sum in (7)
accordingly,
$$+\sum_{k,l\neq i}N_kN_l<\sigma v>_{kl}=+N_ib_i,$$
where $N_ib_i$ is the statistical expected number of the i-th
species produced per unit volume per unit time for a fixed $N_i$.
Note that by nature the number density of species $i,
N_i=N_i(t)$,
is a function of time while the $<\sigma v>_{mn}$ are supposed to
depend only on the temperature but not on the time $t$ and number
densities $N_j$. Then equation (7) implies that
\begin{equation}
\frac{d}{dt}N_i(t)=-(a_i-b_i)N_i(t).
\end{equation}
For equation (9) we have three cases, $c_i=a_i-b_i>0, c_i<0,
c_i=0,$ of which the last case says that $N_i(t)$ does not vary
over time, which means that the forward and reverse reactions
involving species $i$ are in equilibrium. The first two cases
exhibit that either the destruction $(c_i>0)$ of species $i$ or
production $(c_i<0)$ of species $i$ dominates.\\
For the case $c_i>0$ we have
$$\frac{d}{dt}N_i(t)=-c_iN_i(t),$$
and it follows that
\begin{equation}
N_i(t)dt=N_i(0)e^{-c_it}dt,
\end{equation}
where $N_i(0)$ is the number density of species $i$ at time
$t=0$.
If $c_i$ in (10) is a function of time, say $c_i(t)$, then $c_it$
in (10) is to be replaced by $\int dt c_i(t)$. If the arrival
distributions for the other species are Poisson, then $c_i(t)$
will
be of the form $d_it$, where $d_i>0$ independent of $t$. In this
case the exponent in (10) is $\int dt c_i(t)=d_it^2/2.$
Contrarily,
when $c_i$ is $a$ constant, the total number of reactions in the
time interval $0\leq t \leq t_0$ is given by
\begin{equation}
\int^{t_0}_0dt N_i(t)=N_i(0)\int^{t_0}_0dt
e^{-c_it}=\frac{N_i(0)}{c_i}(1-e^{-c_it_0}).
\end{equation}
In (11), $1-e^{-c_it_0}$ is the probability that the lifetime of
species $i$ is $\leq t_0$ when $t$ has the density
\begin{equation}
f(t)=c_ie^{-c_it}, 0\leq t \leq \infty, c_i>0,
\end{equation}
or \begin{equation}
N_i(t)=\frac{N_i(0)}{c_i}f(t).
\end{equation}
When $c_i=c_i(t)=d_it$ then
\begin{equation}
N_i(t)=\left(\frac{\pi}{2d_i}\right)^{1/2}N_i(0)h(t),
\end{equation}
where
\begin{equation}
h(t)=\left(\frac{2d_i}{\pi}\right)^{1/2}e^{-d_it^2/2}, 0\leq t
\leq
\infty, d_i>0.
\end{equation}
The density in (12) will be called the lifetime density for the
destruction of species $i$, with the expected mean lifetime
\begin{equation}
E(t)=\frac{1}{c_i}.
\end{equation}
If the lifetime density is as given in (15) then
\begin{equation}
E(t)=\left(\frac{2}{\pi d_i}\right)^{1/2}.
\end{equation}
{}From (12) and (16) we can make the following observations:\\
(i)\hspace{3em} $c_i$ can be interpreted as a measure of net
destruction, the larger the value of $c_i$ the faster the net
destruction. \\
(ii)\hspace{2em} $\frac{N_i(0)}{c_i}f(t)\Delta t$ can be
interpreted as the amount of net destruction over the small
interval of time $\Delta t$. The faster the net destruction the
shorter the lifetime.\\
(iii)\hspace{1em} The quantity
\begin{equation}
\int^\infty _0 dt \frac{N_i(0)}{c_i}f(t)=\frac{N_i(0)}{c_i}
\end{equation}
can be interpreted as the total net destruction of species $i$
starting with the initial number $N_i(0)$.\\
(iv)\hspace{2em} If the net destruction of species $i$ produces a
species $k$, for example a neutrino, then the number produced is
proportional to $\frac{N_i(0)}{c_i}$.\\
If the lifetime for the production of a species $k$ due to the
net
destruction of species $i$ is denoted by $\tau$, then $\tau$ is a
constant multiple of $t$, say $\tau=\alpha_1t$, where $t$ has the
lifetime density $f(t)$. But the densities of $t$ and
$\alpha_1t\,
(\alpha_1>0)$ belong to the same family of  distributions and
hence
the density of $\tau$ can be written as
\begin{equation}
f(\tau)=\theta_i e^{-\theta_i \tau}, \theta_i >0, \tau>0,
\end{equation}
where $\theta _i=c_i/\alpha_1$ and thus the total production is
$\alpha_1\frac{N_i(0)}{c_i}.$\par
\bigskip
\noindent

\section{Dampening of Reactions: Poisson Arrivals}
Suppose that after a certain period of time of net destruction,
say
$t_0$, a dampening effect starts to slow down the net destruction
of species $i$ with initial number $N_i(0).$ Let this dampening
variable be denoted by $\tau_2$, where $\tau_2$ is again
proportional to the lifetime, say $\alpha_2t$. Then the lifetime
density associated with $\tau_2$ is of the exponential type,
belonging to the same family as in (19). Let $\tau_1$ and
$\tau_2$
be independently acting or statistically independent. Let the
delay
in time for $\tau_2$ to start be $c=\alpha_2t_0$ and let the
densities of $\tau_1$ and $\tau_2$ be denoted by
\begin{equation}
f_j(\tau_j)=\beta_je^{-\beta_j\tau_j}, \tau_j>0, \beta_j>0, j=1,2
\end {equation}
where $\beta_1=\theta_i=c_i/\alpha_1$ of (19) and let
$\beta_2=c_i/\alpha_2.$ Then the net destruction of species $i$
is
proportional to $u=\tau_1-(\tau_2-c)=\tau_1-\tau_2+c$ with the
joint desity of $\tau_1$ and $\tau_2$ given by
\begin{equation}
f(\tau_1,
\tau_2)=\beta_1\beta_2e^{-(\beta_1\tau_1+\beta_2\tau_2)},
\tau_j>0, \beta_j>0, j=1,2
\end{equation}
due to the statistical independence of $\tau_1$ and $\tau_2$. The
density of u, denoted by g(u), is the following (Mathai, 1993)
\begin{equation}
g(u)=\left\{ \begin{array}{cc}
\frac{\beta_1 \beta_2}{\beta_1+\beta_2} & e^{-\beta_1(u-c)},
c\leq
u <\infty \\[0,3cm]
\frac{\beta_1\beta_2}{\beta_1+\beta_2} & e^{\beta_2(u-c)},
-\infty<
u \leq c
\end{array} \right.
\end{equation}
where
$$\frac{\beta_1\beta_2}{\beta_1+\beta_2}=\frac{c_i}{\alpha_1+
\alpha_2},$$
observing that $\beta_1=c_i/\alpha_1$ and $\beta_2=c_i/\alpha_2.$
If the net destruction of species $i$ is exceeding the dampening
rate, then $\beta_1>\beta_2$ and the following Figure 6
illustrates
the behaviour of the density $u$ in (22).\\
Figure 6 shows a non-symmetric Laplacian, slowly rising and
rapidly
falling. At the time $t=t_0$ the net destruction of species $i$
is
given by $\frac{N_i(0)}{c_i}(1-e^{-c_it_0}).$ Then the production
of species, for example neutrinos, as a result of the net
destruction of species $i$, in an instant of time is given by
\begin{equation}
\alpha_1\alpha_2\left(\frac{N_i(0)}{c_i}\right)^2\left(1-e^{-c_it
_0}\right)g(u)du,
\end{equation}
which is a constant multiple of g(u), where g(u) is given in
(22).
Hence the shape of the curve for the net destruction of species
$i$
and the resulting production of species k will be the same as of
g(u) shown in Figure 6. The production of resulting species in a
small intervall of time $\Delta t$ is $Ag(u)\Delta t$ with
$t_0=c/\alpha_2$ starting with a constant initial number $N_i(0)$
of species $i$, where
\begin{equation}
A=\alpha_1\alpha_2\left(\frac{N_i(0)}{c_i}\right)^2\left(1-e^{-
\frac{c_i}{
\alpha_2}c}\right),
\end{equation}
since $\int^{+\infty}_{-\infty}du g(u)=1.$ Here the integration
is
done from $-\infty$ to $+\infty$. Note however, that when $c$ is
large enough the probability for $u$ being negative will be
negligibly small and hence $A$ in (24) is a good approximation to
the total production.\par
We observe in (24) that when $c$ is small, $A$ is small and $A$
is
an increasing function of $c$ as shown in Figure 7.\par
Note that $A$ in (24) is the result of assuming that the initial
number  $N_i(0)$ of species $i$ per unit volume is a constant. If
the species $i$ is arriving to the unit volume according to a
Poisson distribution with parameter $\lambda_i$ (Poisson
arrivals),
then $N_i(0)$ in (24) as well as in the previous formulae is to
be
replaced by its expected value, that is $E[N_i(0)]=\lambda_i$ in
the considered case. In Poisson arrivals one can take the
expected
number to be $\lambda_i=\gamma_it$, where $t$ is the duration of
destruction and $\gamma_i$ is a constant independent of time $t$.
In this case $A$ in (24) becomes
\begin{equation}
A=\frac{\alpha_1\alpha_2}{c_i^2}\gamma_i^2 t^2(1-e^{-c_it_0}),
\end{equation}
where $t_0=c/\alpha_2$ is the time where the dampening effect
starts.\par

\section{Proton-Proton Chain: Branches II and III}\par
The fusion of four protons to produce one helium nucleus in the
pp-chain is accomplished in  at least three different branches in
the chain (Figure 1). This branching results in uncertainties of
the predictions of the $^7$Be- and $^8$B neutrino fluxes in the
Standard Solar Model and needs particular attention in discussing
the results of those solar neutrino experiments which are
detecting
exclusively $^7$Be- and $^8$B neutrinos (Homestake and Kamiokande
experiments in Tables 1,2, and 4). Without any branching in the
pp-chain, the number of all reactions and neutrinos would be
equal,
that means
$$\phi_\nu(pp)=\phi(^8B)=N.$$
As shown in Figure 1, $^3$He can interact with another $^3$He
nucleus to produce right away $^4$He (PPI branch), or $^3$He can
fuse with $^4$He to produce a $^7$Be nucleus and subsequently to
open branches II and III of the pp-chain. The branching ratio r
is
determined by the reaction probabilities $<\sigma v>_{ij}$ and
number densities $N_i$:
\begin{equation}
\frac{r}{1-r}=\frac{<\sigma v>_{34}}{<\sigma
v>_{33}}.\frac{N_4}{N_3},
\end{equation}
where the notations have been explained in the preceding section.
With regard to branches II and III in Figure 1, $^7$Be can
capture
an electron to emit a $^7$Be neutrino, or it can fuse with a
proton
to produce $^8$B which immediately decays and produces a $^8$B
neutrino. The branching ratio $r'$ for PPII and PPIII is
\begin{equation}
\frac{r'}{1-r'}=\frac{<\sigma v>_{17}}{<\sigma
v>_{e7}}.\frac{N_1}{N_e}.
\end{equation}
With (26) and (27) the following relations between the number of
chains to produce $^4$He and the neutrino fluxes produced by the
three branches are established,
\begin{equation}
\phi_\nu(pp)=\frac{N}{2}(2-r), \,\,\,(PPI),
\end{equation}
\begin{equation}
\phi_\nu(^7Be)=\frac{N}{2}r(2-r'), \,\,\,(PPII),
\end{equation}
\begin{equation}
\phi_\nu(^8B)=\frac{N}{2}rr', \,\,\,(PPIII).
\end{equation}
Equations (28) to (30) show the link of the three branches of the
pp-chain which is eventually governed by the reaction
probabilities
$<\sigma v>_{ij}$ and number densities $N_i$ in the system of
kinetic equations in (7) and by the profiles of density and
temperature of the solar model. Basic assumptions for equations
(28) to (30) are that the Sun is in thermal equilibrium which
fixes
the number of chains through (2) and that the nuclei responsible
for neutrino production are in thermal equilibrium with the
ambient
plasma which allows to determine the neutrino fluxes by the
reaction probabilities. For the latter assumption the
characteristic time for significant energy exchange by Coulomb
collisions between reacting species must be orders of magnitude
less than the characteristic time it takes to produce a neutrino
in
the solar interior (Maxwell-Boltzmann reaction rates). These
basic
assumptions still leave the question open on what is relevant for
branching governed by kinetic equations: The time for reducing
the
protons to thermal equilibrium with the ambient plasma ($\approx
10^{-20}$yr) or the lifetime of a proton to undergo a reaction
with
a second proton to produce, among other species, a neutrino
$(\approx 10^{10}$yr)? This question will be addressed in the
following section for three reactions of the branches II and III
of
the pp-chain.

\section {Production - Dampening Mechanism: Laplacian
Behaviour}\par
\medskip
\noindent
Consider three sets of Laplacians of the type given in Figure 6,
one set consisting of one Laplacian with $t_0=\frac{1}{2}(1)$
units
of time, five successive Laplacians with $t_0=\frac{1}{2}(0.2)$
units of time each in the second set, and the third set
consisting
of eight successive Laplacians with $t_0=\frac{1}{2}(0.125)$
units
of time each. Suppose that we consider the Laplacians for a total
arbitrary time interval of $t=1$ unit of time. Let the total
destruction of species $i$ by one Laplacian of set 1, the five
Laplacians of set 2, and the 8 Laplacians of set 3 be denoted by
$A_1, A_2, A_3$ respectively. Then we have from (25)
\begin{eqnarray}
A_1 &=&
\frac{\alpha_1\alpha_2}{c_i^2}\gamma_i^2[1(1)^2]\left(1-e^{-c_i
\frac{1}{2}(1)}
\right),\nonumber \\
A_2 & = &
\frac{\alpha_1\alpha_2}{c_i^2}\gamma_i^2[5(0.2)^2]\left(1-e^{-c_i
\frac{1}{2}(0.2)}\right), \\
A_3 & = &
\frac{\alpha_1\alpha_2}{c_i^2}\gamma_i^2[8(0.125)^2]\left(1-e^{-c
_i\frac{1}{2}(0.125)}\right).\nonumber
\end{eqnarray}
If $c_i$ is large so that $e^{-c_i(.)}$ is negligible, then the
total contributions coming from the three sets are respectively,
\begin{equation}
\frac{A_j}{A_1+A_2+A_3}=0.755,\, 0.15,\, 0.095,\, j=1,2,3
\end{equation}
respectively, that is, 75.5\%, 15\%, and 9.5\% for each of the
three reactions.\par
The Laplacians can be approximated by using triangles in the
following way. From Figure 6 it is noted that the maximum height
of
the Laplacian is at $u=c$ which will then be
\begin{equation}
\frac{\beta_1\beta_2}{\beta_1+\beta_2}=\frac{c_i}{\alpha_1+\alpha
_2}.
\end{equation}
Suppose that $\alpha_2=3\alpha_1$, which will imply that
$\beta_1=\frac{c_i}{\alpha_1}$ and
$\beta_2=\frac{c_i}{3\alpha_1}$,
which in turn means that the net destruction rate is three times
the dampening rate. Suppose that $\beta_1=\sqrt{3}b$, where $b>0$
is a constant and $t_0=\frac{3}{4}b.$ Then the maximum hight of
the
Laplacian in Figure 6 is
\begin{equation}
\frac{\beta_1\beta_2}{\beta_1+\beta_2}=\frac{c_i}{\alpha_1+\alpha
_2}=
\frac{\sqrt{3}b}{4}.
\end{equation}
In this case the Laplacian approximates to the following triangle
shown in Figure 8.\par
\noindent
If we take three sets of triangles shown in Figure 8, where the
first set consists only of one triangle with $b=1$ time unit, the
second set contains 5 successive triangles with $b=0.2$ time
units
each, and in the third set there are 8 successive triangles with
$b=0.125$ units each, and if the total areas of these three sets
of
triangles are denoted by $A_1, A_2,$ and $A_3$ similiar to (31),
then the respective areas are in the proportion 75.5, 15, and 9.5
percent respectively (see Table 5).\par
\begin{tabular}{|l|c|l|}\hline
Total area & Contribution to the & $^{37}$Ar production rate of\\
& 9 year cycle & the Homestake experiment \\
& &(Table 3)\\ \hline
1 triangle & 75.5 \% & $^8$B contributes 77.5\%\\
Reaction 1 & &\\ \hline
5 triangles & 15.0\% &$^7$Be contributes 15\%\\
Reaction 2 & &\\ \hline
8 triangles & 9.5\% &\\
Reaction 3 & &\\ \hline
\end{tabular}\par
\vspace{0.5cm}
Table 5.\\
\noindent
Triangles as the ones shown in Figure 8 have been used in the
following graph, where in Figure 9, $\alpha$ and $\beta$ denote
the
shifts in the starting point of the set of 5 and of the set of 8
triangles, respectively. The starting point of the one big
triangle
has been chosen as $t=0$.\par
\bigskip
\noindent

\section {Conclusion}
\medskip
\noindent
The time variation of the argon production in the Homestake
experiment which is ascribed to be  produced by solar neutrinos
can
be explained as follows. The original Homestake data, but more
distinctively the five-point moving average of the data, seem to
show cycles of approximately nine years duration. The reactions
of
the PPII and PPIII branches of the proton-proton chain are
producing neutrinos through the $^7Be$- and $^8B$- reactions. If
one assumes that a dampening mechanism operates for three
reactions
of the PPII and PPIII branches as discussed above, the variations
of the argon production in the Homestake experiment over time can
be explained on purely statistical arguments based on lifetimes
and
their ratios for the three reactions. For these nuclear
reactions,
destruction and dampening may work opposite to each other. If the

destruction rate is approximately three times the dampening rate,
if the destruction is $\sqrt{3}b$ for some $b>0$, and if the
dampening effect starts $\frac{3}{4}b$ time units from the
starting
time $t=0$, then the time variation cycles seen in the argon
production in the Homestake experiment can be reproduced by
considering a scenario of three sets of reactions of the PPII and
PPIII branches of the proton-proton chain, one set with $b=1$
unit
of time, say 9 years, the second set consisting of 5 successive
reactions with $b=0.2$ time units each, and the third set
consisting of 8 successive reactions with $b=0.125$ time units
each.\par
\bigskip
\begin{center}
References
\end{center}
Abdurashitov, J.N. et al.: 1994, Phys. Lett. \underline{B328},
234.\par
\smallskip
\noindent
Anselmann, P. et al.: 1994, Phys. Lett. \underline{B327},
377.\par
\smallskip
\noindent
Bahcall, J.N. and Pinsonneault, M.H.: 1992, Rev. Mod. Phys.
\underline{64}, 885.\par
\smallskip
\noindent
Castellani, V. et al.: 1994, Phys. Rev. \underline{D50},
4749.\par
\smallskip
\noindent
Davis Jr., R.: 1993, in Y. Suzuki and K. Nakamura (eds.),
'Frontiers of\par Neutrino Astrophysics', Universal Academy
Press,
Inc., Tokyo.\par
\smallskip
\noindent
Haubold, H.J. and Gerth, E.: 1990, Solar Physics \underline{127},
347.\par
\smallskip
\noindent
Haubold, H.J. and Mathai, A.M.: 1994, in H.J. Haubold and L.I.
Onuora\par
(eds.), 'Basic Space Science', AIP Conference Proceedings Vol.
320,\par
American Institute of Physics, New York.\par
\smallskip
\noindent
Kiko, J.: 1995, The GALLEX solar neutrino experiment at the
GranSasso\par Underground Laboratory; these Proceedings.\par
\smallskip
\noindent
Mathai, A.M. and Haubold, H.J.: 1988, Modern Problems in Nuclear
and\par
Neutrino Astrophysics, Adademie-Verlag, Berlin.\par
\smallskip
\noindent
Mathai, A.M.: 1993, Canad. J. Statist. \underline{21}, 277.\par
\smallskip
\noindent
Nakamura, K.: 1993, Nucl. Phys. Suppl. \underline{B31}, 105.\par
\clearpage
\noindent
\underline{Table captions:}\par
\begin{tabbing}
Table 1: \=Predictions of the Standard Solar Model for the
Kamiokande,\\             \>Homestake, and GranSasso/Baksan
experiments (Bahcall and\\          \>Pinsonneault, 1992).\\
[0.3cm]
Table 2: \> The four currently operating solar neutrino
experiments.\\ [0.3cm]
Table 3: \> Predicted capture rates in SNU from various flux
components of the\\
         \> pp-chain and CNO cycle for the Homstake and
GranSasso/Baksan\\
         \>experiments. The uncertainties are the total
theoretical
range,\\
         \>$\sim 3\sigma$ (Bahcall and Pinsonneault, 1992).\\
[0.3cm]
Table 4: \>Comparison of the detected rates of the four solar
neutrino          experiments\\
         \>with the predicted rates of the Standard Solar Model
(Bahcall and\\
         \>Pinsonneault, 1992). The Kamiokande flux is in units
of
$10^6\nu cm^{-2}s^{-1}$,\\
         \>while the Homestake and GranSasso/Baksan rates are in
SNU.\\[0.3cm]
Table 5: \>For the destruction-dampening mechanism considered
here,
the area\\
         \>of the triangle governing reaction 1, the combined
areas
of the\\          \>5 triangles for reaction 2, and the combined
areas of the \\
         \>8 triangles of reaction 3, are proportional to the
combination\\          \>of three sources to the total capture
rate
of the Homestake experiment.
\end{tabbing}
\clearpage
\noindent
\underline{Figure captions:}\par
\noindent
\begin{tabbing}
Fig. 1: \=The proton-proton chain and its three different
branches
to accomplish\\
         \> to formation of $^4He$.\\[0.3cm]
Fig. 2:  \>The neutrino production as a function of the
dimensionless distance\\
         \>variable $x=R/R_\odot$ in the Standard Solar Model of
Bahcall and\\
         \>Pinsonneault (1992).\\[0.3cm]
Fig. 3:  \>The argon-production rate detected by the Homestake
experiment from\\
         \>1970.8 to 1991.6 (Davis, 1993).\\[0.3cm]
Fig. 4:  \>The five-point moving average of the argon-production
rate data as\\
         \>shown in Figure 3.\\ [0.3cm]
Fig. 5:  \>Power spectrum of the argon-production rate data in
Figure 3 obtained\\
         \> by Fourier analysis.\\[0.3cm]
Fig. 6:  \> A non-symmetric Laplacian, slowly rising and rapidly
falling.\\
         \> The function $g(u)$ is the density of $u$, describing
the destruction-\\
         \>dampening mechanism for nuclear reactions involving
species i.\\[0.3cm]
Fig. 7:  \> The behaviour of the function A in (24) denoting the
total destruction of\\
         \> species i. \\[0.3cm]
Fig. 8:  \>The non-symmetric Laplacian shown in Figure 6 can be
approximated by \\                   \>a non-symmetrical
triangle.\\ [0.3cm]
Fig. 9:  \>The destruction-dampening mechanism in (25) and (26),
where the Laplacians\\
         \> has been approximated by triangles, can reproduce the
time variation of\\
         \> the original argon-production rate within the range
of
the error bars\\
         \> attached to each run.
\end{tabbing}
\end{document}